\documentclass[preprint]{aa}
\usepackage{txfonts}
\usepackage{graphicx}
\usepackage{hyperref}

\def\asec{\ifmmode ^{\prime\prime}\else$^{\prime\prime}$\fi}

\def\it{\sl}
\def\degs{\ifmmode ^{\circ}\else$^{\circ}$\fi}
\def\amin{\ifmmode ^{\prime}\else$^{\prime}$\fi}
\def\asec{\ifmmode ^{\prime\prime}\else$^{\prime\prime}$\fi}

\def\degs{\ifmmode ^{\circ}\else$^{\circ}$\fi}
\def\amin{\ifmmode ^{\prime}\else$^{\prime}$\fi}

\def\eqalign#1{\null\,\vcenter{\openup1\jot \m@th
   \ialign{\strut\hfil$\displaystyle{##}$&$\displaystyle{{}##}$\hfil
   \crcr#1\crcr}}\,}
\begin{document}


\authorrunning{Zharikov, S., et al.}
\titlerunning{Cyclic brightening in the short-period CV SDSS 0804 }
\title{Cyclic brightening in the short-period WZ~Sge-type cataclysmic variable SDSS~J080434.20$+$510349.2
}
\author{S.V. Zharikov\inst{1}
\and G. H. Tovmassian\inst{1}
\and V. V. Neustroev\inst{2}
\and R.~Michel\inst{1}
\and C.~Zurita\inst{1}
\and J.~Echevarr\'{i}a\inst{3}
\and I.F.~Bikmaev\inst{4}
\and E.P.~Pavlenko\inst{5}
\and   Young- Beom Jeon\inst{6}
\and G.G.~Valyavin\inst{6}
\and A.~Aviles\inst{1}}
\institute{
Observatorio Astron\'{o}mico Nacional SPM, 
Instituto de Astronom\'{i}a, Universidad Nacional Aut\'{o}noma de M\'{e}xico,
Ensenada, BC, M\'{e}xico 
\and
Centre for Astronomy, National University of Ireland, Galway, Newcastle Rd., Galway, Ireland 
\and
Instituto de Astronom\'{i}a, Universidad Nacional Aut\'{o}noma de M\'{e}xico,  D.F., M\'{e}xico
 \and
 Department of Astronomy, Kazan State University, Russia
\and
Crimean Astrophysical Observatory, Nauchny, Ukraine
\and
Bohyunsan Optical Astronomy Observatory, South Korea
}

\offprints{S. Zharikov,\\
\email{zhar@astrosen.unam.mx}}

\date{Received --- ????, accepted --- ????}

\abstract{} {We 
observed a new 
cataclysmic variable (CV) SDSS J080434.20$+$510349.2  to
study the origin of  long-term variability  found in its light
curve. } 
{Multi-longitude, time-resolved, photometric
observations were 
acquired
to analyze this uncommon behavior, which has been 
 found 
in two newly discovered CVs. }
{ This study of {SDSS J080434.20$+$510349.2} 
concerns primarily   the
understanding of
the nature of the observed, double-humped, light curve and
its relation to a
cyclic brightening that
occurs during
quiescence. The observations were obtained
early in 2007, when the object was at about $V\sim17.1$, about 0.4
mag brighter than the pre-outburst magnitude.
The light curve shows
a sinusoidal variability with an amplitude of about 0.07 mag
and a periodicity of 42.48 min,
which is half of the
orbital period of the system. 
We  observed in addition two
 ``mini-outbursts" of the system of
up to 0.6 mag, which have  a duration of about 4 days each.  The ``mini-outburst" has
a symmetric profile  and is repeated in approximately every 32 days. Subsequent
monitoring of the system shows a cyclical behavior
of such ``mini-outbursts" 
with a similar recurrence period. The origin
of the double-humped light curve and the periodic
brightening  is discussed in the light of the evolutionary
state of SDSS J080434.20$+$510349.2. }
{} 
{  }

\maketitle

\section{Introduction}
SDSS J080434.20$+$510349.2 ({hereafter} {SDSS\,0804})
was 
identified as a faint ($B\sim 18$ mag), short-period (P$_{\mathrm{orb}}=85\pm3$ min) cataclysmic variable by
Szkody et al.  (\cite{szkody06}). These authors reported 
that the optical spectrum of SDSS~0804, in quiescence, shows a blue
continuum with broad absorption lines from a white dwarf, which surround
the double-peaked Balmer emission lines formed in an accretion disk.
The spectrum is similar to the spectra of WZ Sge-type systems.

On 2006 March 4, Pavlenko et al. (\cite{pavlenko06})
observed this star during a
super-outburst with $V_{max} = 12.8$~mag.  At the end of
the super-outburst,  eleven  echoes
took place, with an interval of 2.6 days. Such post-outburst activity has been   observed in only  a handful of CVs, all of which are  of WZ Sge type.  Echoes are therefore considered to be 
 a characteristic  property of WZ~Sge stars.
Inspection of archive plates from Sonneberg (1923-2006) and Odessa
(1968-1993) reveals only one previous outburst
($\sim$12.5 mag), which occurred in 1979 (Pavlenko et al.
\cite{pavlenko06}). Szkody et al. (\cite{szkody06}) reported that
the light curve of {SDSS 0804} showed a 42.5 minute periodic
variability with an amplitude of $\sim$0.05~mag, which is half the
spectroscopic orbital period. These double-humped light curves
are observed often  in the eary stages of an outburst
in WZ Sge systems, and on rare occasions  in quiescence. Imada et al.
(\cite{imada06})  proposed to include the presence of
double-peaked light curves in short-period CVs as an additional
criterion for a WZ Sge-type classification. Thus, SDSS\,0804 exhibits
all the necessary attributes to be classified as a classical WZ Sge
object: a short orbital period, infrequent and large-amplitude
super-outbursts succeeded by echo outbursts, a double-humped light
curve, and other
features such as strong emission lines surrounded by broad
absorption and long-lasting super-humps during  a super-outburst.

In addition to the ``standard'' set of WZ Sge features, Szkody et al. (\cite{szkody06}) detected 
a rapid rise in brightness of the system by 0.5 mag , 
at the same time as the amplitude of the 42.5 minute variation increased to about 0.2 mag (hereinafter named as ``\emph{brightening}'').
A similar behavior - the large increase in brightness together with the increase in amplitude of the modulation - was
first discovered by Zharikov et al. (\cite{zhar06}) in another
short period CV SDSS J123813.73-033933.0
(hereafter abbreviated as {SDSS\,1238}), where such
brightenings are cyclic.
Both objects have also a 
similar spectral appearance
in quiescence.

\begin{table*}[t]
\begin{center}
\caption{Log of time-resolved observations of SDSS J080434.20$+$510349.2}
\begin{tabular}{llllccc}
\hline\hline
 Date        & HJD Start+ & Telescope& Band & Exp.Time& Duration\\
Photometry & 2454000 &   &    &                 Num. of Integrations   &                            &                \\ \hline
12 Dec. 2006 & 82.888   &1.5m/SPM &  R    & 180s$\times$101                     & 4.20h  \\
13 Dec. 2006 &  83.794  &1.5m/SPM &  R    & 180s$\times$101                     & 5.77h  \\
14 Dec. 2006 &  84.804  &1.5m/SPM &  R    &  180s$\times$119                    & 5.76h  \\
15 Dec. 2006 &  85.802  &1.5m/SPM &  R    & 120s$\times$129                        & 5.81h  \\ \hline
6  Jan. 2007 & 107.317  &1.5m/RTT150 & V    & 120s$\times$133                     & 8.32h  \\
7  Jan. 2007 & 108.276  &1.5m/RTT150 &  V    & 120s$\times$128                     & 9.31h  \\
8  Jan. 2007 &  109.278 &1.5m/RTT150 &  V    & 120s$\times$143                     & 9.48h  \\
9  Jan. 2007 & 110.263  &1.5m/RTT150 &  V    &120s$\times$135                      & 10.03h  \\
10  Jan. 2007 & 111.468  &1.5m/RTT150 &  V    &120s$\times$135                      & 5.11h  \\
11  Jan. 2007 &  112.325 &1.5m/RTT150 &  V    &120s$\times$135                      & 8.57h  \\ \hline
10  Jan. 2007 & 111.428  &0.8m/IAC80 &  WL    &120s$\times$135                     & 7.44h  \\
14  Jan. 2007 & 115.390  &0.8m/IAC80 &  WL  &120s$\times$135                      & 9.19h  \\
15  Jan. 2007 &  116.369 &0.8m/IAC80 &  WL &120s$\times$135                      & 9.67h  \\ \hline
15 Jan. 2007 &  116.745  &0.84m/SPM &  V    & 180s$\times$101                     & 7.34h  \\
16 Jan. 2007 &  117.617  &0.84m/SPM &  V    &  180s$\times$119                    & 6.52h  \\
17 Jan. 2007 &  118.624 &0.84m/SPM &  V    & 120s$\times$129                        & 10.53h  \\
20 Jan. 2007 &  121.629  &0.84m/SPM &  V    & 180s$\times$101                     & 8.89h  \\
21 Jan. 2007 &  122.881  &0.84m/SPM &  V    & 180s$\times$101                     & 3.55h  \\
22 Jan. 2007 &  123.623  &0.84m/SPM &  V    & 180s$\times$101                     & 9.20h  \\
23 Jan. 2007 &  124.659  &0.84m/SPM &  V    &  180s$\times$119                    & 8.81h  \\
24 Jan. 2007 &  125.617  & 0.84m/SPM &  V    & 120s$\times$129                    & 7.90h  \\ \hline
25 Jan. 2007 &  126.671  &1.5m/SPM &  V    & 180s$\times$101                     & 7.56h  \\
26 Jan. 2007 &  127.626  &1.5m/SPM &  V    & 180s$\times$101                     & 8.62h  \\
27 Jan. 2007 &  128.623  &1.5m/SPM &  V    &  180s$\times$119                    & 8.66h  \\
28 Jan. 2007 &  129.672  & 1.5m/SPM &  V    & 120s$\times$129                    & 5.90h  \\ \hline
\end{tabular}
\label{tab1}
\end{center}
\end{table*}

Interested by the similarity between the systems, 
 we conducted a new
time-resolved photometric study of SDSS\,0804 
to establish the reasons behind their common nature, understand the origin
of the cyclic brightening and its relation to the amplitude of the double-humped light curve. In Sect.\ref{Obs},  we  describe our observations and data reduction.
The data analysis and the results are presented  in Sect.\ref{DatAn}, while a general discussion  is given  in Sect.\ref{Discus}.

\begin{figure}[t]
 \setlength{\unitlength}{1mm}
 \resizebox{12.cm}{!}{
 \begin{picture}(60,43)(0,0)
 \put (3,0){\includegraphics[width=40mm, bb= 0 0  800 850, clip=]{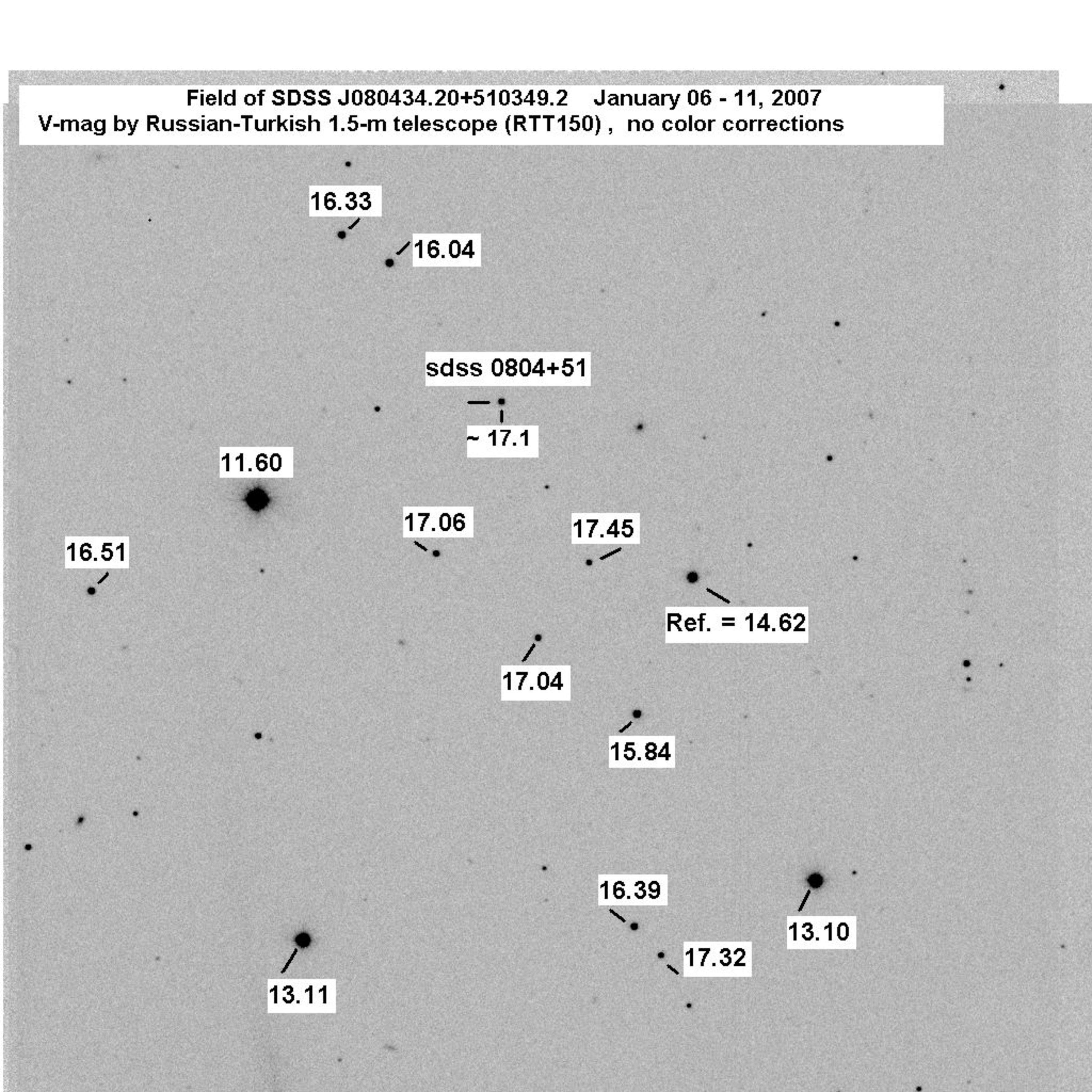}}
 \end{picture}}
 \caption{The field of SDSS\,0804 observed using the RTT150 telescope.
  The north is at the top of the image and the east is at the left.  The image size is $\sim 6.5\times 6.5$ arcmin.
 The object and the secondary standard stars are indicated.
 The V-band magnitudes of the secondary standard stars are marked.}
 \label{fig1}
 \end{figure}

 \begin{figure*}[t]
 \setlength{\unitlength}{1mm}
 \resizebox{12.cm}{!}{
 \begin{picture}(110,85)(0,0)
 \put (0,0){\includegraphics[width=160mm, bb=00 250 600 600, clip=]{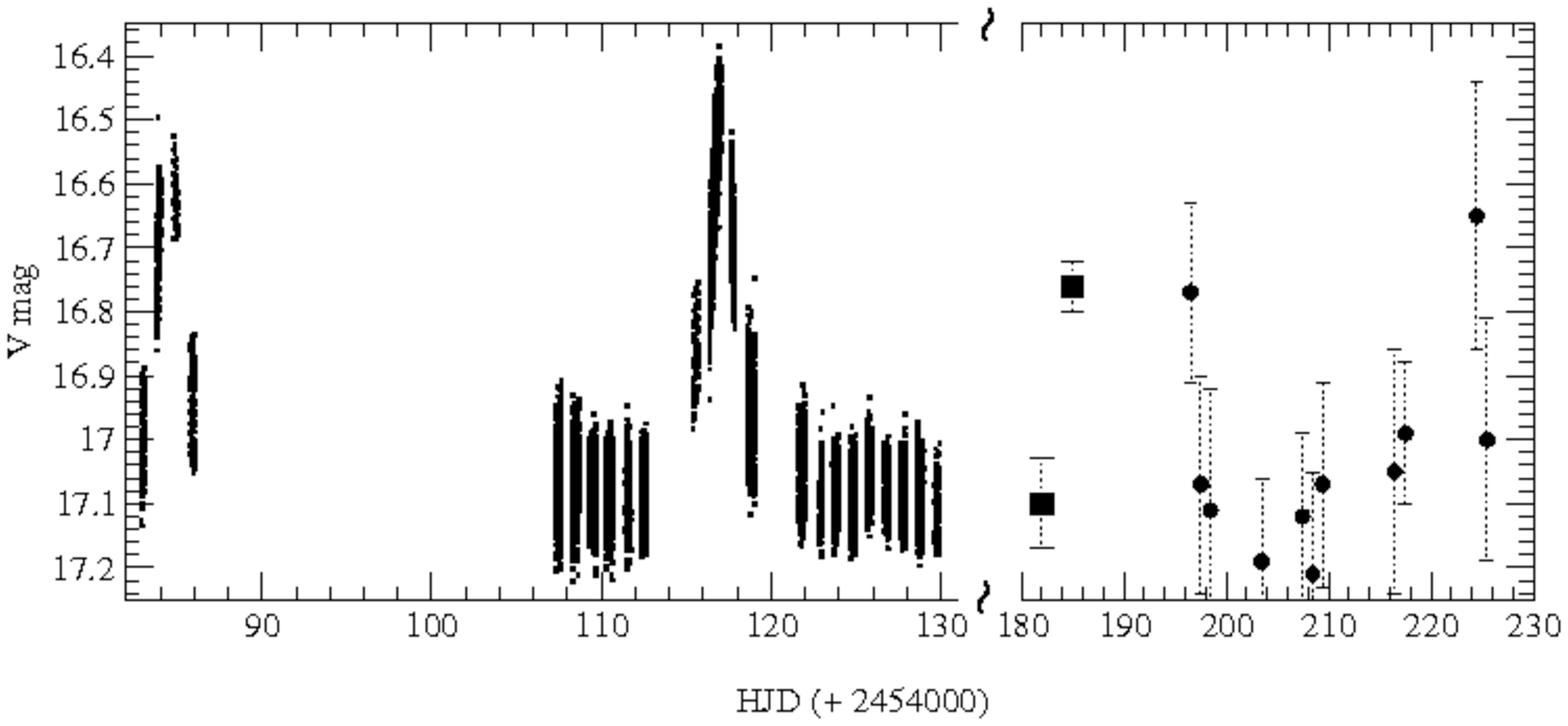}}
 \end{picture}}
 \caption{  The composite light curve of SDSS\, 0804 of data acquired throughout the campaign. The log of observations in the period  HJD 24540882-24541131 is given in the Table.1.
The monitoring of the system in the period HJD 24541180-24541230 is represented by the 2.1m telescope BOAO (full squares) and the 0.4m telescope of the Imbusch observatory (full circles) data.  }
 \label{fig2}
 \end{figure*}

\section{Observations and data reduction}
\label{Obs} Taking into account the long duration of the
\emph{brightening}  and the uncertainty in  a  \emph{brightening}
cycle period of SDSS\,0804, we  planned and executed multi-longitude
observations of this object. Time-resolved photometry  of SDSS\,0804
was obtained using direct CCD image mode
at several facilities:
the 1.5m and 0.84m telescopes at the Observatorio Astron\'omico
Nacional at San Pedro M\'artir
in Mexico; the 1.5m Russian-Turkish telescope
at the TUBITAK National Observatory (TUG) in Turkey; the 0.8m IAC80
telescope at the Observatorio del Teide in the
Canary Islands, Spain; the 2.1m telescope at the Bohyunsan Optical Astronomy Observatory (BOAO) in South Korea; 
and the 0.4m telescope at the Imbusch observatory in Galway, Ireland.
The log of time-resolved
observations is presented in Table~\ref{tab1}. Several field
stars as well as Landolt photometric stars were also observed.

Data reduction
was performed
using both ESO-MIDAS and IRAF software. The images were
bias-corrected and
flat-fielded before aperture photometry was carried out. The errors
of the differential CCD photometry
were calculated
from the dispersion of the magnitude of the comparison
stars.
The dispersion
ranged from 0.01 to 0.05 mag, during the
observational
period HJD 24540882-24541131. The errors
during HJD 24541180-24541230 were 0.05-0.1 mag for  BOAO data, and
0.15-0.2 mag
for  data obtained
using the 0.4m Imbush telescope.
Calibration of
the field stars, observed in the Johnson V-band,
was
obtained from the Landolt
standards, and thus they became
secondary standard stars.
Their corresponding magnitudes are indicated in
Fig.\ref{fig1}. A residual uncertainty in their absolute
calibration may reach  $\sim0.1$ magnitude because of the absence of
color-index information. The magnitudes
in the R-band were derived using  the V and R magnitudes of
the reference star marked in Fig.\ref{fig1},
using the USNO A2.0 catalogue
(Fig.\ref{fig1}). The data obtained without filter (White Light:
marked  WL in Table.\ref{tab1}) were  transformed 
to the V-band.
The light curve of the entire set of observations is presented in Fig. \ref{fig2}.

\section{Data analysis}
\label{DatAn}

\begin{figure*}
 \setlength{\unitlength}{1mm}
 \resizebox{9.cm}{!}{
 \begin{picture}(60,100)(0,0)
 \put (0,0){\includegraphics[width=120mm, clip=]{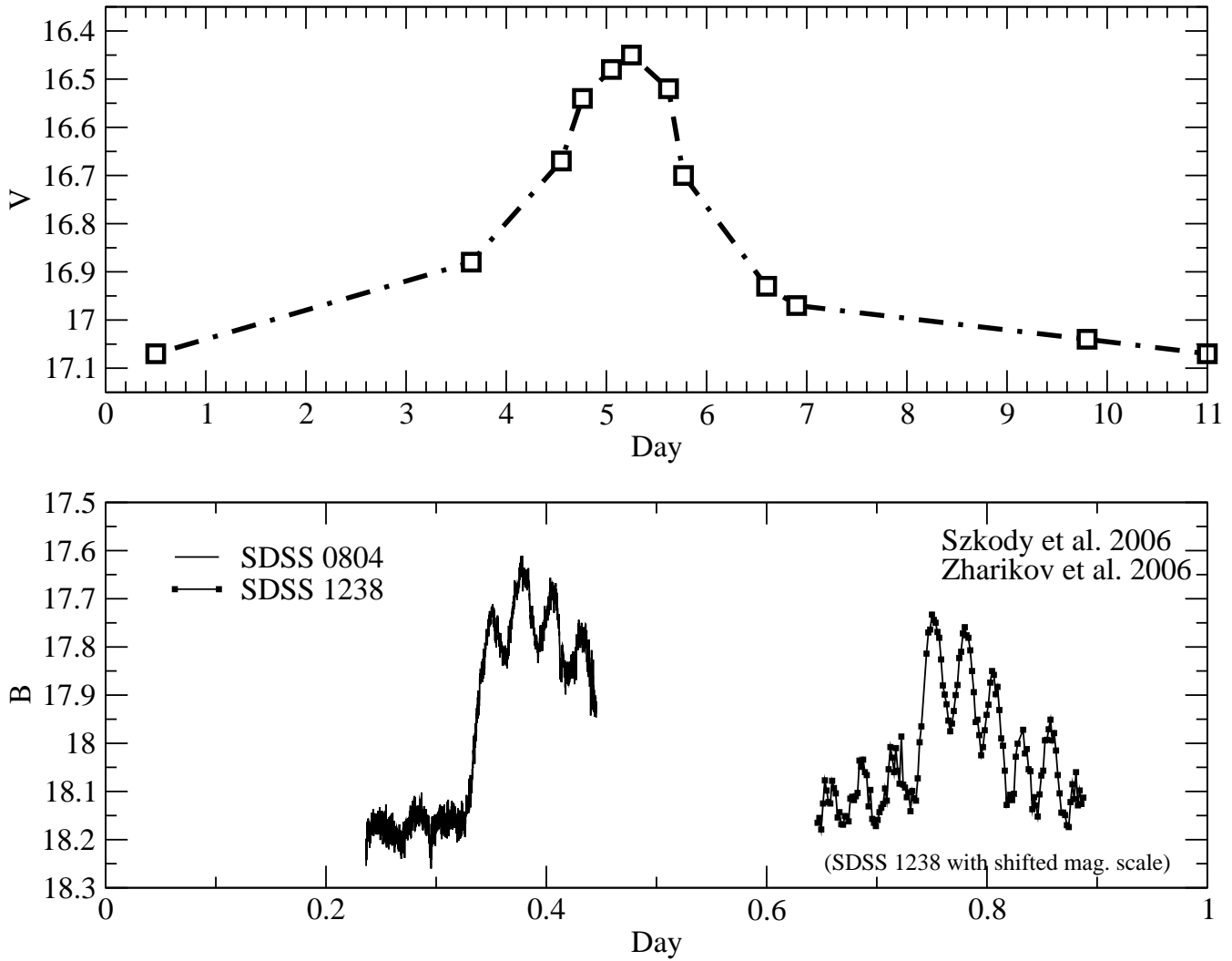}}
 \end{picture}}
 \caption{ The average, smoothed, composite light curve of
 the \emph{mini-outbursts}  (upper panel).
 Examples of
 \emph{brightness} events in
 the SDSS\, 0804 and SDSS\,1238 systems,
 accompanied by a change of the amplitude
 with half-orbital period variability  (bottom panel). }
 \label{fig5}
 \end{figure*}

Frequent  \emph{brightenings} (on the timescale of a fraction of a day) were expected in the system from comparison of the behavior  of SDSS\,0804 (Szkody et al. \cite{szkody06}) with the light curve of SDSS\,1238 (Zharikov et al. \cite{zhar06}).  
 We show examples of  \emph{brightenings} of both objects side by side in the bottom panel of\  Fig.\ref{fig5}, on similar  time and magnitude scales. The \emph{brightening} events for SDSS\,0804 and for SDSS\,1238 have an almost identical behavior.  
 The quiescent state is interrupted by a sudden and fast rise of brightness  
 during a time corresponding to  half the orbital period, with a simultaneous increase in the amplitude of the double-humped variation.  For SDSS\,1238, the brightness increase lasts only $\sim 3$-4 hours and repeats  itself cyclically about every 8-12 hours. The \emph{brightenings} of SDSS\,0804 last for about a similar time 
 but  there is no information 
 on how frequently they  occur 
 prior to  super-outburst.

{ We found repetitive brightness increases in the new observations of SDSS\,0804,  although  their  behavior was  
 different.  Firstly, we  note that the object at the time of our observations  had a brightness of   $V~\sim~17.1$ \,mag, which is, about 0.4 mag brighter than in the quiescent state before the 2006 super-outburst (Pavlenko et al. \cite{pavlenko06}).} Earlier in 2005 (see Szkody et al. \cite{szkody06}), the brightness of the object was  estimated to be  $V\geq 17.5$ mag, judging from the B-band photometry and the $(B-V)\sim0.15$ color index calculated from the SDSS spectrum\footnote{http://www.sdss.org}. 

Secondly, the object  exhibits only two incidents of a brightness increase during the observing period corresponding to HJD 2454082-2454131, defined here as \emph {mini-outbursts} to differentiate them from \emph{brightenings}. The amplitudes of the \emph{mini-outbursts} are about 0.6 mag and are similar to the amplitudes of the \emph{brightenings}.  The \emph{mini-outbursts},  however,  last approximately 4 days, based on comprehensive monitoring 
of two events. A composite profile of  all \emph{mini-outbursts} is presented in the top panel of Fig.\ref{fig5}. Please note, that the timescales of the upper and bottom panels are different. 
The \emph{brightenings} last only $\sim0.2$ days.
 The use of the term \emph{mini-outburst} is  appropriate also 
  because 
 these events do not resemble
dwarf nova outbursts: 
their amplitude is too small 
for an outburst, i.e. the total energy release is significantly smaller than usually produced in an outburst as a result of thermal instability of the accretion disk. The object probably shows two more \emph{mini-outbursts}, as can be seen in the complete light curve of SDSS\,0804 presented in Fig.\ref{fig2}. The time between the first two \emph{mini-outbursts} is 32 days.

Thirdly and most importantly, we found that the object shows a double-humped light curve with constant amplitude, during all of  the time that the object was observed. We do not detect any variation in the amplitude of  the  double humps with respect to luminosity. The brightness variation, referred  to as 
a \emph{mini-outburst},
develops slowly during 2-3 days, reaches a similar  amplitude as that observed during the \emph{brightening}
observed by Szkody et al. (\cite{szkody06}), but shows almost a symmetrical profile. The amplitude of the double-hump variation remains unchanged
throughout the entire \emph{mini-outburst} and equals  the pre and post-\emph{mini-outburst} value.

\begin{figure*}
 \setlength{\unitlength}{1mm}
 \resizebox{9.cm}{!}{
 \begin{picture}(60,60)(0,0)
 \put (0,0){\includegraphics[width=190mm, bb=0 150 900 450, clip=]{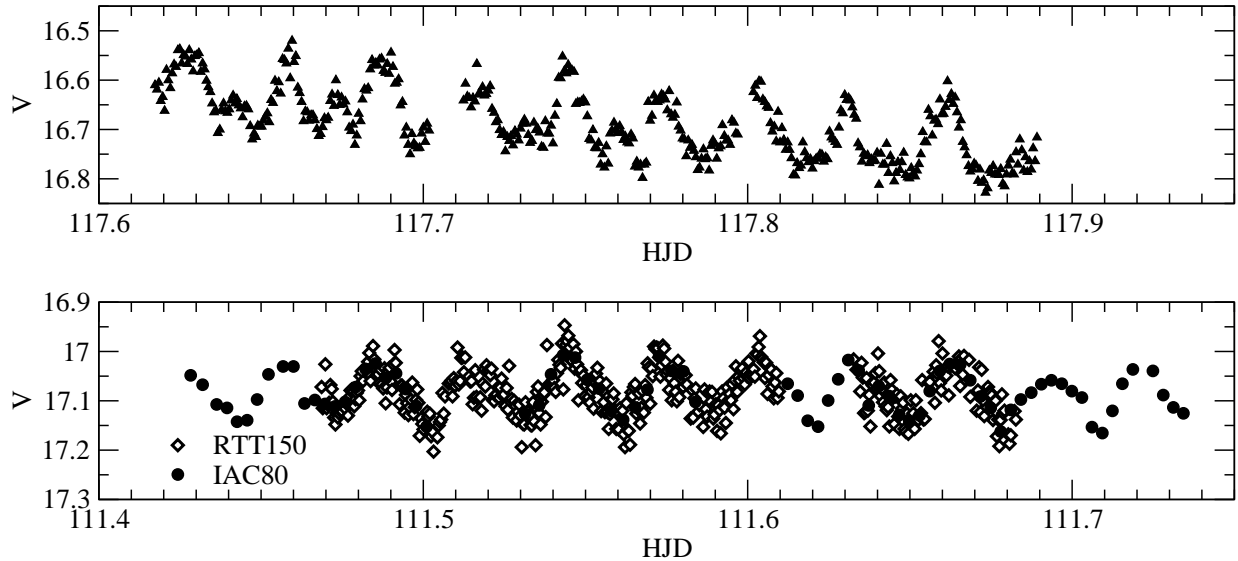}}
 \end{picture}}
 \caption{ Fragments of a light curve  during the ``mini-outburst" and ``repose" states (up and low respectively).}
 \label{fig4}
 \end{figure*}

\begin{figure}
 \setlength{\unitlength}{1mm}
 \resizebox{9.cm}{!}{
 \begin{picture}(60,45)(0,0)
 \put (0,0){\includegraphics[width=73mm, bb= 50 150 800 800, clip=]{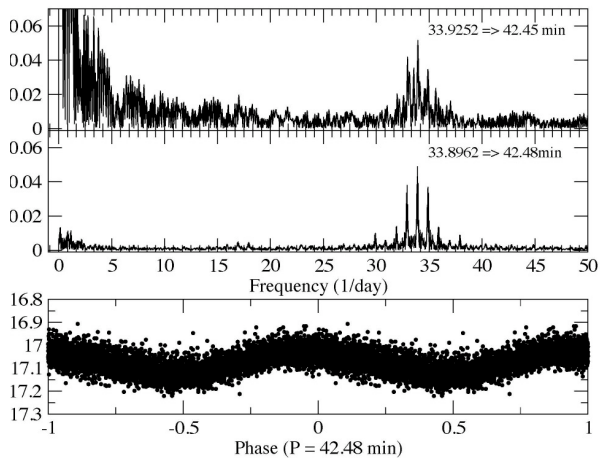}}
 \end{picture}}
 \caption{ The power spectrums obtained for both states: \emph{mini-outburst} (top) and \emph{repose} state (below).
The light curve of the \emph{repose} state folded by the period of
$P_\mathrm{phot}=42.48(2)$ min $=1/2 P_{\mathrm{orb}}$ (bottom
panel). }
 \label{fig3}
 \end{figure}

To complete 
the time analysis, we separated  our time-resolved observation data  into  two distinct categories -
\emph{repose} and \emph{mini-outburst}, which both occurred when the system was mainly in quiescence.
In a state of \emph{repose}, the object flickers around
$V\sim17.1$ (see the lower panel in Fig.\ref{fig4}), while the mini-outburst corresponds to a brightness increase  in the light curve,  where the brightness of the object reaches 16.5 mag at maximum 
%
(see the upper panels in Fig.\ref{fig5} and Fig.\ref{fig4}). The data acquired during the state of \emph{repose}
were analyzed for periodicities using the Discrete Fourier
Transform code (Deeming \cite{Deeming}). The power spectrum of the
\emph{repose} data is presented  in Fig.\ref{fig3}  (middle panel).
The  peak corresponding to the maximum power is located
at $P_\mathrm{phot}=42.48(2)$min,
which corresponds
to half
the orbital period of the system. The $\sim$0.07
mag variability has a sinusoidal shape, as can be seen in the lower
panel of Fig.\ref{fig3}, where the data
are folded by the $P_\mathrm{phot}=42.48(2)$ min
period. The character of the light curve, before and after the
mini-outburst, is completely identical. The period and phase of the
 periodic variations are preserved throughout the \emph{mini-outburst} state.
The power spectrum of the data during \emph{mini-outbursts} (an example of the \emph{mini-outbursts} data is
presented in the upper panel of the Fig.\ref{fig4}) shows a similar
peak in frequency as that for   the double-hump period, but,
in this case, it is contaminated
by the profile of the \emph{mini-outbursts} (Fig.\ref{fig3}, upper panel).

We
 assume that $P_\mathrm{orb}=2\times
P_\mathrm{phot}=0.05900$d,  is
the true orbital period of SDSS\,0804. This value is
within the error range of the spectroscopic orbital period
$P_\mathrm{orb}^{sp}=0.0592(4)$d, derived  by Pavlenko et al.
(\cite{pavlenko07}). Using
our estimated  P$_\mathrm {orb}$ value, we  calculate
the system mass
ratio
to be $q\approx0.05$, based on its super-hump period
$P_\mathrm{sh}=0.059713(7)$d (Pavlenko et al. \cite{pavlenko06}), and
the $\varepsilon = 0.18q+0.29q^2$ relation between a period
excess
$\varepsilon\equiv(P_\mathrm{sh}-P_\mathrm{orb})/P_\mathrm{orb}$ and
a mass
ratio $q\equiv M_\mathrm{2}/M_\mathrm{1}$ (Patterson et al.
\cite{Patt05}). If this empirical relation holds for extremely short periods, then
the mass of the secondary
cannot exceed
0.08$M_\odot$.

 \section{Discussion}

The cataclysmic variable SDSS\,0804,
since its discovery, has been proposed as a WZ~Sge-type
candidate, based on its short orbital period and spectral/photometric
characteristics. 
Further evidence to support this
classification was  the occurrence and morphology of its 2006 super-outburst and consequent
echoes (Pavlenko et al. \cite{pavlenko06}). The echoes,  similar 
to WZ Sge (Patterson et al. \cite{patterson02}) and EG Cnc (Kato et
al. \cite{kato04} and references therein), are considered as an exclusive property of the  WZ Sge class among dwarf novae.
Super-humps with the period of 0.059713~d were also detected during the super-outburst, which led  to 
a low mass ratio  estimate of $q\sim0.05$.
During our observations taken about one year after its super-outburst, SDSS\,0804 remained brighter, by about 0.4 mag, than it was before the March 2006 event.
 A similar increase in the quiescence
level after the super-outburst, was observed  in another WZ Sge-type
system Al~Com  (Nogami et al. \cite{Nogami}).

In addition to  classical WZ~Sge properties, SDSS\,0804 
exhibited a variability that was almost identical to
that  observed for SDSS\,1238. First,  
there was 
the persistent 
double-humped light curve in quiescence, which had however 
variable amplitude. 
Then, there were  the cyclic 
 luminosity  increases of only a half magnitude.

\begin{figure}
 \setlength{\unitlength}{1mm}
 \resizebox{8.cm}{!}{
 \begin{picture}(60,48)(0,0)
 \put (0,0){\includegraphics[width=60mm, clip=]{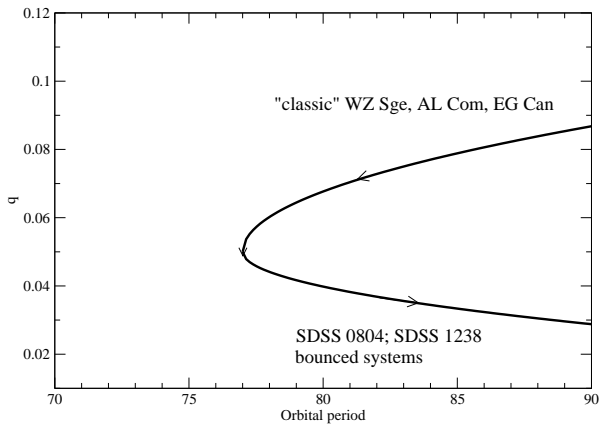}}
 \end{picture}}
 \caption{ The schematic evolution state of ``classical" WZ-Sge system and ``non-classical" SDSS 0804 and SDSS1238. }
 \label{fig6}
 \end{figure}

Various models (see Patterson et al. \cite{patterson02}  and  Imada
et al \cite{imada06} and ref. therein) have been proposed to explain
the double-humped light curve\textbf{s} in WZ Sge systems, among
which  the 2:1 resonance (Lin \& Papaloizou \cite{linpap}, Osaki \&
Meyer \cite{osaki}, Kunze \& Speith \cite{kunze}) in  systems with
mass ratio $q\le0.1$
is favoured. If
this 2:1 resonance is responsible for the double-humped light curves, then we have to account for the difference between ``classical" WZ Sge-type systems (WZ Sge, AL\,Com,
EG\,Cnc) and the newly SDSS-discovered objects (i.e. SDSS\,0804 and
SDSS\,1238), and explain why they undergo cyclical brightenings during
quiescence.
According to the 2:1 resonance model the rim of
the disk expands and reaches the 2:1 resonance region during the
super-outburst  in ``classical" WZ Sge systems.
If SDSS\,0804 and SDSS\,1238 contain less massive
secondaries than ``classical" WZ Sge systems,  it is possible that the radius of the accretion disk in these systems is continually reaching of 2:1 resonance radius.
Less massive secondaries put
 SDSS\,0804 and SDSS\,1238 as
  ``period bounce"  
systems, i.e close binaries which  have
reached the period limit $\sim$77 min boundary and have turned
around (Barker \& Kolb \cite{Barker}), as opposed  to ``classical"
WZ~Sge systems, 
which are still evolving
towards an orbital period minimum (see. Fig.\ref{fig6}).  
Steeghs et al. \cite{Steeghs}  
determined the mass ratio $M_2/M_1$ for the components of WZ\,Sge itself  to be  $0.075 <  q < 0.101$. 
Their  inferred donor mass $M_2 = 0.078\pm0.06 M_\odot$ corresponds to an L2-type star and according to
Knigge (\cite{knigge2006}), the system  still evolves toward its period minimum.
Using a large range of masses for the white dwarf in  SDSS\,0804 $ 0.6M_\odot < M_1 < 1.4M_\odot$ and our estimate of $q=0.05$  we calculate that the  secondary mass is in the range  $0.03 M_\odot < M_2 < 0.07 M_\odot$, which makes  it more likely to be a postÐperiod minimum system.

The observations of SDSS\,1238 
 show cyclical or quasi-periodic brightenings 
with a sudden increase 
 in 
amplitude of  the  double-hump curve (Zharikov et al. \cite{zhar06}).
The same is probably true for  the pre-outburt behavior of SDSS\,0804.
The cyclical nature of  the \emph{brightenings}
suggests that the mass-transfer rate varies
cyclically too. 
If this is the case, even a small increase in the mass
transfer rate of a system will cause an expansion of
the accretion disk,
 with a rapid increase in the
brightness of the system and a long extended tail in the
decay phase (Ichikawa \& Osaki \cite{ichikawa92}). In combination 
with the disk-size increase, the two-armed spiral dissipation
pattern will form and emerge as a double-humped light curve (Kunze
\& Speith \cite{kunze}). 

After the 2006 super-outburst,
the behavior of SDSS\,0804 in
quiescence has qualitatively changed. Firstly, nine months after 
super-outburst the system has still
not descended to its pre-outburst quiescent level,
but remains about 60 percent brighter than it was before.
Secondly, during the entire duration  of our observational campaign, the
system displayed
a double-humped light curve of approximately similar amplitude.
Thirdly, the cyclical brightenings have changed significantly. The
timescale of this change exceeds significantly that of 
the brightenings observed for SDSS\,1238 (Zharikov et al.
\cite{zhar06}), and that observed for SDSS\,0804 by
Szkody et al. (\cite{szkody06}).
The recurrence time, compared to that of SDSS\,1238, is incompatibly
longer. The shape is different and the brightening, or mini-outburst 
to differentiate it from conventional brightenings, is similar to that of
 a  normal
outburst in SU UMa systems, but it has significantly  lower amplitude,
when compared to the typical amplitude in DN systems ranging from 2 to 6 magnitudes,  or is 
as   a  ``stunted" outburst such as that  observed in some  nova-like cataclysmic
variables (Honeycutt \cite{honeycutt}). This is a new phenomenon
that has not been observed before in other  WZ Sge-like systems in
quiescence. Finally and most importantly, the amplitude of the
double-humped variation does not  depend on the mini-outburst occurrence.

Another mechanism is required
to explain the \emph{mini-outbursts} in
addition to the double-humped light curve and the
brightenings
prior to the super-outburst.
We find that such  a mechanism could be an irradiation of the accretion
disk as a result of  the super-outburst. The small amplitude
post-eruption outbursts with an amplitude of only $\sim0.5$
mag   were
predicted by Hameury et al. (\cite{Hameury}).
These authors modeled
the time-dependent behavior of irradiated accretion disks in dwarf
novae and post novae following an outburst. The
contribution of irradiation by the white dwarf
to the inner parts of the accretion disk was found to cause small
outbursts, followed immediately
by normal outbursts, or even a super-outburst.
Since such outbursts
had been not  observed before,
Hameury et al. (\cite{Hameury}) concluded, that either the inner
disk
 was evaporated or the efficiency of the irradiating flux
from the white dwarf
was  lower than  expected.
According to their models, these small outbursts, 
or \emph{mini-outbursts}, as we call them  to distinguish
them from normal outbursts, start, however, as inside-out outbursts, which die
out before
they reach half  the radius of the accretion
disk, unable to propagate across the entire disk. This  leaves
the outer parts of the disk intact, where the 2:1 resonance occurs.
In the case of SDSS\,0804,
the internal parts of the disk have probably not been
destroyed during the super-outburst and
hence, it is possible to observe the prolonged effect of
irradiation in a dwarf nova directly.
Furthermore, the irradiation of the disk might be the same
mechanism that produces the echos  appearing
after
the super-outburst in SDSS\,0804 and some other WZ Sge systems.
 Because of the peculiar mass ratio of a period-bounce system, the mass-transfer 
 rates and, the accretion disk size, we observed the effect of  irradiation  for a
long period of time.


 \label{Discus}
 \section{Conclusions}
 \label{conclude}

We observed SDSS\,0804 almost a year after it underwent  a
super-outburst. 
The system exhibits all  the attributes of a
WZ\,Sge-type system and, in addition,  shows low-amplitude cyclical
mini-outburst activity,  
which causes them to become 
brighter than during the pre-outburst quiescent
level. We identify these  \emph{mini-outbursts} as the small inside-out outbursts
predicted by Hameury et al. (\cite{Hameury}) as a result of
an irradiation of a disk by a powerful super-outburst.  The  \emph{mini-outbursts} 
 differ from the  \emph{brightenings}  
  observed previously in
 SDSS\,0804 and SDSS\,1238. 
The  \emph{brightenings}  have similar amplitude as the \emph{mini-outbursts}   but 
show a different temporal behavior and therefore a smaller energy output. 
We suggest that  variable mass transfer produces  the \emph{brightenings}, and directly influences the   2:1 resonance
effect, which determines  the amplitude of  the double-hump light curve.
 On the other hand,  the \emph{mini-outbursts}  are of a sporadic nature as a result of  irradiation of the accretion disk 
and  are not related to  the amplitude of the double humps. 
 We argue  that both of these
 CVs have probably evolved beyond the period limit, and
hence, are members of long sought, elusive bounced-back systems, and
therefore differ from other WZ\,Sge systems.

The new time-resolved spectral observations  of these system with
high signal/noise ratio obtained during quiescence, would be useful help us 
understand  the accretion-disk structure changes that correspond to the
re-brightening phenomena. In addition, the numeric  simulation of
the accretion disk dynamic in 2:1 resonance can  help us to
understand the dynamics of the evolution of spiral-armed structures in
accretion disks and their observational properties.

\begin{acknowledgements}
This  work  was partially supported   by  PAPIIT  IN101506 and CONACYT 48493 projects.
VN acknowledges support of IRCSET under their basic research programme and
the support of the HEA funded CosmoGrid project.
We wish to thank Prof. Mike Redfern for help with the observations at the
Imbusch observatory.
We  thank the anonymous  referee, for  
comments that led to an improved presentation of the paper.
\end{acknowledgements}

\end{document}